\documentclass[prb,showpacs,twocolumn]{revtex4}
\usepackage{graphicx}
\usepackage{bm}
\usepackage{pifont}
\usepackage{amsmath}
\begin{document}
\title{Orbital characters of three-dimensional Fermi surfaces in Eu$_{2-x}$Sr$_{x}$NiO$_{4}$
as probed by soft-x-ray angle-resolved photoemission spectroscopy}
\author{
M. Uchida$^{1}$, K. Ishizaka$^1$, P. Hansmann$^{2}$, X. Yang$^3$, M. Sakano$^{1}$, J. Miyawaki$^{4}$, R. Arita$^1$, Y. Kaneko$^5$,
Y. Takata$^{4}$, M. Oura$^{4}$, A. Toschi$^2$, K. Held$^{2}$, A. Chainani$^{4}$, O. K. Andersen$^{6}$, S. Shin$^{4,7}$, and Y. Tokura$^{1,5,8}$
}
\affiliation{
$^1$Department of Applied Physics, University of Tokyo, Tokyo 113-8656, Japan\\
$^2$Institut for Solid State Physics, Vienna University of Technology, 1040 Vienna, Austria\\
$^3$School of Materials Science and Engineering, Nanyang Technological University, Singapore 639798, Singapore\\
$^4$RIKEN SPring-8 Center, Hyogo 679-5148, Japan\\
$^5$Multiferroics Project, ERATO, Japan Science and Technology Agency (JST), Tokyo 113-8656, Japan\\
$^6$Max-Planck-Institut f\"{u}r Festk\"{o}rperforschung, D-70569 Stuttgart, Germany\\
$^7$Institute for Solid State Physics (ISSP), University of Tokyo, Chiba 277-8561, Japan\\
$^8$Cross-Correlated Materials Research Group (CMRG) and Correlated Electron Research Group (CERG), RIKEN Advanced Science Institute (ASI), Wako 351-0198, Japan
}
\begin{abstract}
The three-dimensional Fermi surface structure of hole-doped metallic layered nickelate Eu$_{2-x}$Sr$_x$NiO$_4$ ($x=1.1$),
an important counterpart to the isostructural superconducting cuprate La$_{2-x}$Sr$_x$CuO$_4$,
is investigated by energy-dependent soft-x-ray angle-resolved photoemission spectroscopy.
In addition to a large cylindrical hole Fermi surface analogous to the cuprates,
we observe a $\Gamma$-centered $3z^2-r^2$-derived small electron pocket.
This finding demonstrates that in the layered nickelate the $3z^2-r^2$ band resides close to the $x^2-y^2$ one in energy.
The resultant multi-band feature with varying orbital character as revealed
may strongly work against the emergence of the high-temperature superconductivity.
\end{abstract}
\pacs{71.18.+y, 79.60.-i, 74.25.Jb}
\maketitle
High-temperature superconductivity in the cuprate family
has been a central issue of interests in the field of strongly correlated electron systems.
While an essential role of band-filling control (charge doping) in the CuO$_2$ plane with a quantum spin ($S=1/2$) state has been clarified,
the bottom line of the material conditions for the high-temperature superconductivity remains elusive.
In this context, designing analogs of the layered cuprates with other transition metal oxides
and comparing their electronic states should help understanding of uniqueness of the superconducting cuprates.
The electronic structures of several layered antiferromagnets with $S=1/2$,
such as $3d^1$ vanadates \cite{Vanalog1, Vanalog2, Vanalog3} and $3d^7$ nickelates, \cite{Nianalog1, osmtLNO1, osmtLNO2}
have been mainly theoretically examined from this angle.
Recently, the Mott-insulating $5d^5$ layered iridates with an effective moment $J=1/2$
have also attracted much attention as potential host materials. \cite{IrhalfJ1, IrhalfJ2, Iranalog1, Iranalog2}
A comprehensive experimental study on such analogs can place strong constraints on possible microscopic mechanisms of the superconductivity,
although it has achieved only limited success so far.

Single-layer nickelate $R_{2-x}$Sr$_x$NiO$_4$ ($R$: rare earth elements), isostructural to superconducting La$_{2-x}$Sr$_x$CuO$_4$ (LSCO), 
has attracted attention for long as one of the most important counterpart materials to the superconducting cuprates.
\cite{Nianalog1, NiJ, Nipoly1, Nipoly2, Nithin, Nistripe1, Nistripe2, Nistripe3, NiCB1, NiCB2, NiFS}
The parent compound ($x=0$) is an antiferromagnetic insulator with $T_\mathrm{N} \sim 330$ K, \cite{NiJ}
and with hole-doping this system also shows a quasi-two-dimensional antiferromagnetic insulator-metal transition at about $x=1$. \cite{Nipoly1, Nipoly2, Nithin}
In the low-doping region, it has been found that the doped holes spontaneously form diagonal-stripe ($x\sim 1/3$) \cite{Nistripe1, Nistripe2, Nistripe3} and checkerboard ($x\sim 1/2$) \cite{NiCB1, NiCB2} charge orders in the NiO$_2$ plane.
While anomalous metallic behaviors represented by strongly temperature-dependent Hall resistivity and mostly-incoherent charge dynamics
have recently been reported on the verge of the Mott transition ($x\sim1.0$-1.1) \cite{Nithin, NiFS} in close analogy to the cuprate case,
superconductivity has not been found in the nickelates so far. \cite{Nipoly1, Nipoly2}
Considering that this nickelate system shares many other conditions which have been thought to be prerequisites for the high-temperature superconductivity,
such as quasi two-dimensionality, a proximate quantum spin insulating state, and charge and antiferromagnetic-spin correlations,
one important point to be considered is the orbital configuration of $R_{2-x}$Sr$_x$NiO$_4$ around $x=1$ ($3d^7$).
Especially if the $d_{x^2-y^2}$ energy level were lower than the $d_{3z^2-r^2}$ one,
holes would be doped into the single $x^2-y^2$ band in the barely metallic state above $x=1$,
realizing an electronic state similar to the superconducting cuprates.
This situation has been theoretically predicted for a LaNiO$_3$/LaAlO$_3$ superlattice system, \cite{osmtLNO2}
which is anticipated to host the high-temperature superconductivity as well.
Recent laser angle-resolved photoemission spectroscopy (ARPES) of Eu$_{2-x}$Sr$_x$NiO$_4$ (ESNO)
has indeed clarified a fraction of the large hole Fermi surface (FS) analogous to the cuprates. \cite{NiFS}
Local-density approximation (LDA) calculation, on the other hand, shows that the $3z^2-r^2$ band exists in the vicinity of the $x^2-y^2$ one
and that the FSs consist of the large hole-like cylinder and an additional $3z^2-r^2$-derived small electron-like sphere, as shown in Fig. 1(a).
FS structure and its orbital character should be closely related to the emergence of the high-temperature superconductivity, \cite{Cureview}
and thus investigation of the complete three-dimensional (3D) FS in the layered nickelate is
of crucial importance for revealing the criteria which separate the non-superconducting nickelates and the superconducting cuprates.
Here we report $h\nu$-dependent soft-x-ray (SX)-ARPES experiment on ESNO at $x=1.1$
and discuss the contrasting FS structures with focus on the multi-orbital character.

\begin{figure}
\begin{center}
\includegraphics*[width=8.6cm]{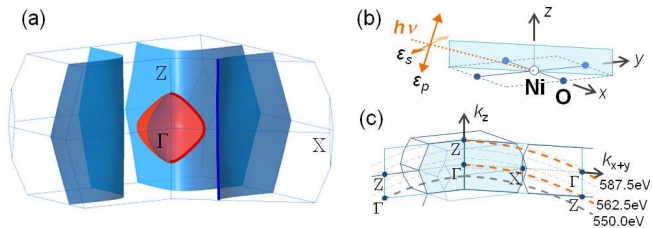}
\caption{(Color online).
(a) Calculated 3D FSs of ESNO ($x=1.1$) in a Brillouin zone of the body-centered tetragonal structure.
(b) Experimental configuration of a NiO$_2$ plane, light polarization vectors and the mirror plane ($1\bar{1} 0$), under the grazing light incidence.
(c) Momentum regions of the SX-ARPES measurement over the 3D Brillouin zones.
The momentum cuts at the typical incident photon energies correspond to
the arcs shown with thick dashed curves, passing through some high-symmetry points ($\Gamma$ and Z).
We used an inner potential and a work function of 15 and 5 eV, respectively, to make the periodic $k_z$-dependence consistent.
}
\label{fig1}
\end{center}
\end{figure}

Single crystals of ESNO ($x=1.1$) were grown by the floating-zone method in a high-pressure oxygen atmosphere ($p_{\textrm{O}_{2}}=60$ atm).
SX-ARPES experiments with excitation energies of $h\nu= 510-590$ eV
were carried out with a hemispherical electron analyzer,
Gammadata-Scienta SES2002, at undulator beam line BL17SU of SPring-8. \cite{BL}
The SX-ARPES endstation employs grazing incidence geometry to increase the photoelectron yield,
as shown in the schematic setup configuration in Fig. 1(b).
Its geometry also ensures that the x-ray photon momentum correction can be simply taken into account
($\Delta k_{x+y} \sim 0.2 \Gamma$-X at $h\nu = 550$ eV),
with negligible modification of the $k_{z}$ momentum.
The momentum cuts traced by the specific photon energies mainly used are shown in Fig. 1(c).
The ARPES measurements were performed in a high vacuum of $1 \times 10^{-10}$ torr, at 45 K.
The energy resolution was set to 150 meV.
Along with circularly polarized x-rays, to observe all the orbital components,
linearly polarized x-rays were also utilized to selectively extract the $d$-orbital character.
The degree of the linear polarization was about 80\%.
The LDA calculation was performed with the linear and $N$th-order muffin-tin orbital methods, \cite{LMTO}
as described in detail elsewhere. \cite{NiFS}

Figure 2 represents the overall 3D FSs of ESNO ($x=1.1$) as observed by SX-ARPES.
In an example of the in-plane FS shown in Fig. 2(a), measured at $h\nu =550.0$ eV,
($\pi, \pi$)-centered large hole FS contours are clearly visible with the parallel straight segments around the ($\pi , 0$) regions.
The image plot along the momentum cut A shown in Fig. 2(b) indicates
that the band dispersion around the ($\pi , 0$) point has a van Hove singularity (VHS) at about 0.2 eV below $E_{\mathrm{F}}$,
and thus it ensures that the FS indeed remains hole-like despite its large volume.
Hereafter, we use the terminologies ``nodal" and ``antinodal" for the ($0, 0$)-($\pi , \pi$) and ($0, 0$)-($\pi , 0$) directions of the observed large hole FS,
respectively, to comprehensively discuss the common electronic structures between the layered nickelates and cuprates.

\begin{figure}
\begin{center}
\includegraphics*[width=8.6cm]{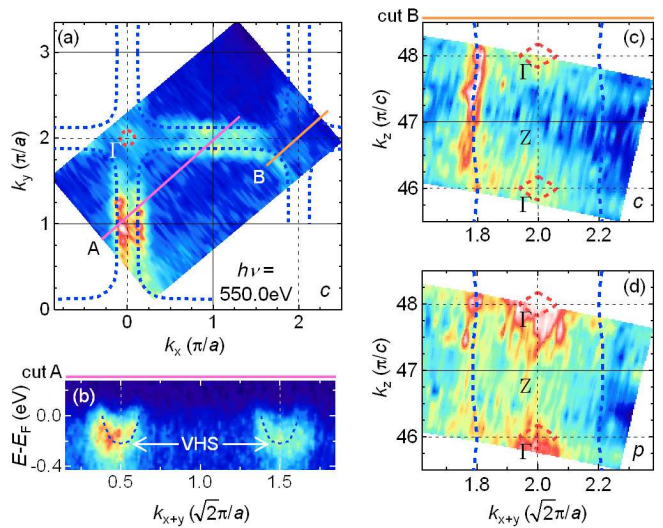}
\caption{(Color online).
3D FS topology of ESNO ($x=1.1$).
(a) FS map in the $k_{x}$-$k_{y}$ plane taken with 550.0 eV circularly polarized light.
(b) Image of the band dispersion corresponding to the momentum cut A shown in (a).
(c) and (d) FS maps in the $k_{x+y}$-$k_{z}$ plane including the momentum cut B and the $\Gamma$-Z line,
taken using circular and linear $p$ polarization light, respectively.
The photon energy range of $h\nu=535.0$-587.5 eV is used.
The momentum-space maps are obtained by integrating the photoemission intensity within an $E_{\mathrm{F}} \pm 75$ meV window.
The thick dashed lines represent the $k_{\mathrm{F}}$ positions or the energy dispersions obtained from the MDC peak positions.
}
\label{fig2}
\end{center}
\end{figure}

In Fig. 2(c) we show the $k_z$-dependent FS map containing the nodal momentum cut B, as obtained using the circular polarization light.
It shows a rather 2D component at $k_{x+y}\sim 1.8 \sqrt {2} \pi /a$ with small warping along the $k_z$ direction,
which corresponds to the aforementioned ($\pi, \pi$)-centered hole-like FS.
In addition, it presents weak but distinct intensity distribution around the $\Gamma$ point, suggesting the existence of another 3D FS.
Figure 2(d) gives the FS map in the same region, measured by adopting the linear $p$ polarization.
The $\Gamma$-centered spectral weight is more clearly discerned in comparison with the quasi-2D hole FS,
indicating that the aforementioned small FS exists around the $\Gamma$ point besides the large cylindrical hole FS.

From now on, we proceed to the detailed electronic structures forming the two FSs.
Figures 3(a) and 3(b) present the in-plane FSs at $k_{z}=0$ and $\pi /c$ planes, measured at $h\nu =587.5$ and 562.5 eV, respectively.
The ARPES intensity from the small FS can be seen around the $\Gamma$ point, while no intensity around the Z point.
Images of the band dispersions forming the FSs are shown in Figs. 3(c)-(f).
To distinguish the bands consisting of $x^2-y^2$ or $3z^2-r^2$ orbital character, we utilized the $s$ and $p$ polarization.
In our experimental geometry (Fig. 1(b)), $s$ ($p$) polarization has odd (even) parity with respect to the mirror plane,
which contains all the photoemission vectors.
The initial state $x^2-y^2$ ($3z^2-r^2$) wave function has also odd (even) parity with respect to the mirror plane,
and then the matrix element argument \cite{selectionrule} leads to a simple selection rule
that a band dispersion with $x^2-y^2$ ($3z^2-r^2$) character appears only in the $s$ ($p$) configuration.
As shown in Fig. 3(c), at $k_{z}=0$ plane, the dispersion of the large hole FS is clearly seen
but that of the small FS is not observed with the $s$ polarization.
In the case of the $p$ polarization shown in Fig. 3(d), by contrast, the hole band dispersion almost disappears
while a shallow electron pocket dispersion clearly appears.
It indicates that the large hole sheet and small electron pocket have the $x^2-y^2$ and $3z^2-r^2$ characters along the nodal direction, respectively.
Since hybridization between the two $e_g$ orbitals is forbidden along the nodal (0, 0)-($\pi , \pi$) line, \cite{OKA}
the discernible hole band with $p$ polarization (Fig. 3(d)) is not due to the orbital hybridization but may arise from the limited degree ($\sim 80$\%) of the linear polarization.
At $k_{z}=\pi /c$ plane, on the other hand, only the $x^2-y^2$-derived hole band dispersion is observed, as shown in Figs. 3(e) and 3(f).
These contrastive behaviors can be more clearly discerned in the corresponding momentum distribution curves (MDCs) shown in Figs. 3(g)-3(j).
Well-defined MDC peaks (Fig. 3(h)) represent the clear dispersion of the electron pocket with a depth of about 0.1 eV.
Our polarization-dependent ARPES results thus fully demonstrate that
the $3z^2-r^2$-derived electron FS indeed exists around the $\Gamma$ point, as predicted by the LDA band calculation.
As for the hole band dispersion (Figs. 3(c) and 3(e)), in addition,
kink-like structures are clearly seen on a high-energy scale ($\sim 0.2$-0.4 eV) for both $k_{z}=0$ and $\pi /c$ planes.
This deviation from the bare band dispersion, termed the high-energy kink,
has been observed at a comparable energy ubiquitously in the superconducting cuprates. \cite{firstgiant, hierar}
It indicates that the layered nickelate indeed shares the high-energy physics behind the anomaly, 
such as a coupling to bosons \cite{giantboson} or renormalization by the strong electron correlation, \cite{hierar, giantNCCO}
with the case of the cuprates. 

\begin{figure}
\begin{center}
\includegraphics*[width=8.6cm]{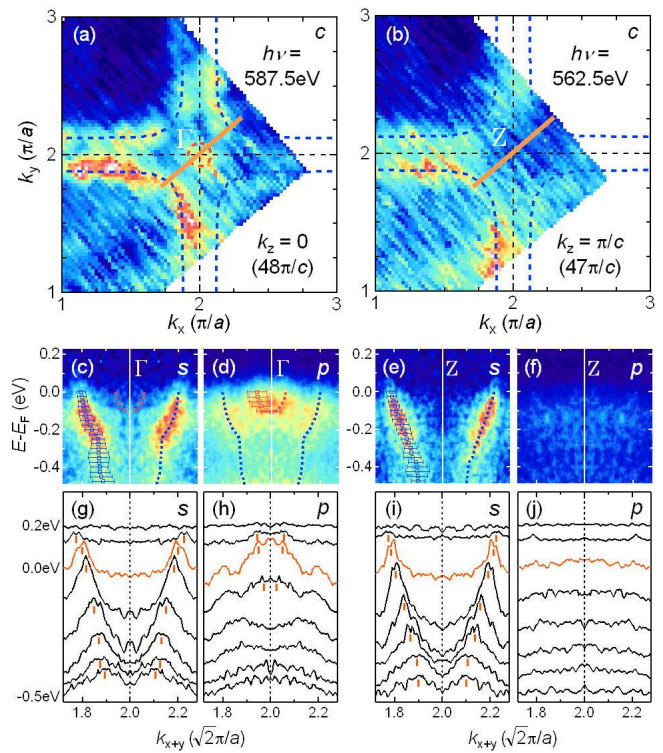}
\caption{(Color online).
In-plane FSs measured with (a) 587.5 and (b) 562.5 eV circularly polarized light.
They correspond to those in $k_{z}=0$ and $\pi /c$ planes, respectively.
The intensity maps are taken using the same integration window of $E_{\mathrm{F}} \pm 75$ meV as in Fig. 2.
The thick dashed lines show the $k_{\mathrm{F}}$ positions determined by the MDCs at $E_{\mathrm{F}}$.
(c)-(f) Image plots of band dispersions with $s$ and $p$ polarization at the momentum cuts shown in (a) and (b).
For clarity, data have been symmetrized at the $\Gamma$ and Z points.
The dashed lines show the energy dispersions experimentally determined by following the MDC peak positions like the open squares with error bars.
(g)-(j) Stacks of the corresponding MDC profiles plotted for every 0.1 eV.
} 
\label{fig3}
\end{center}
\end{figure}

Among the fundamental electronic structure of the metallic layered nickelate as revealed by the present SX-ARPES,
one remarkable difference from the cuprate family is the existence of the small but definite $3z^2-r^2$-derived electron pocket at the zone center.
Such a multi-band nature of the FS as derived from the orbital degeneracy, not observed in the superconducting cuprates,
may strongly inhibit the $d$-wave superconductivity.
To be noted furthermore is the overall orbital character of the large hole FS,
considering that the $3z^2-r^2$ band across the Fermi energy can also cause the orbital mixing away from the nodal (0, 0)-($\pi , \pi$) direction. \cite{OKA}
Figure 4(a) illustrates the in-plane momentum dependence of the relative ($3z^2-r^2$)/($x^2-y^2$) character obtained by the LDA calculation.
The $\Gamma$-centered electron FS mainly consists of the $3z^2-r^2$ component.
For the hole FS, except at the nodal point with the genuine $x^2-y^2$ character as confirmed by the polarization-dependent ARPES,
the two $e_g$-orbital components are nearly comparable over a wide momentum-region including from off-nodal to antinodal directions.
To compare to the case of the isostructural layered cuprate,
we show in Fig. 4(b) the corresponding calculation result for LSCO at optimal doping ($x=0.15$).
Also in this superconducting cuprate, the $3z^2-r^2$ orbital is rather strongly hybridized at the antinodal ($\pi , 0$) region.
From nodal to off-nodal regions, however, the $x^2-y^2$ contribution is widely dominant in contrast to the case of ESNO.
In various superconducting cuprates including LSCO, some systematic experiments have indicated a possibility that
only the near-nodal quasiparticles contribute to the $d$-wave superconductivity while the antinodal ones remain incoherent. \cite{pgBitwo, Ramantwo, LSCOtwo}
In the light of this scenario, the comparison between ESNO and LSCO suggests that the near-nodal orbital character would be of great significance
for the emergence of the superconductivity.

\begin{figure}
\begin{center}
\includegraphics*[width=8.2cm]{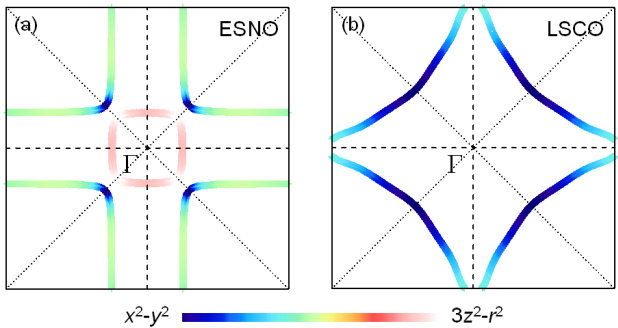}
\caption{(Color online).
Momentum-space variation of the calculated ($x^2-y^2$)/($3z^2-r^2$) character of the FSs
at $k_{z}=0$ plane in (a) ESNO ($x=1.1$) and (b) LSCO ($x=0.15$).
Focusing on the large hole-like FS, the two $e_g$ orbitals are strongly hybridized at the antinodal region in both the systems.
As for the near-nodal region, on the other hand, the relative characters are nearly comparable in ESNO except at the nodal point,
while the $x^2-y^2$ character remains highly dominant in LSCO.
Similar momentum dependence was obtained also at other $k_{z}$ cuts (not shown).
}
\label{fig4}
\end{center}
\end{figure}

We have thus clarified the overall differences of the FS and its orbital character between the layered nickelates and cuprates.
Our findings indicate that in the nickelate system a suite of the FS changes,
the appearance of the small electron pocket, and the non-negligible orbital mixing especially near the off-nodal direction,
results from the $e_g$-band energy level difference.
While the present approach from the non-superconducting analog suggests that
multi-band features with varying orbital character as revealed may work against the emergence of the high-temperature superconductivity,
further control of the $e_g$-band level from the presented level, such as by chemical substitutions and fabrications of multilayer/heterostructure,
will lead to the development of nickelate-based superconductors.

This work was partly supported by a Grant-in-Aid for Scientific Research (Grant No. 20340086) from JSPS,
and by Funding Program for World-Leading Innovative R\&D on Science and Technology (FIRST Program).
M.U. acknowledges support by a Grant-in-Aid for the JSPS Fellowship program (Grant No. 21-5941).
The synchrotron radiation experiments were performed at BL17SU in SPring-8 with the approval of RIKEN (Proposal No. 20110073).
%\newpage

%

\begin{thebibliography}{100}
\bibitem{Vanalog1} W. E. Pickett, D. Singh, D. A. Papaconstantopoulos, H. Krakauer, M. Cyrot, and F. Cyrot-Lackmann, Physica C \textbf{162-164,} 1433 (1989). 
\bibitem{Vanalog2} Y. Imai, I. Solovyev, and M. Imada, Phys. Rev. Lett. \textbf{95,} 176405 (2005).
\bibitem{Vanalog3} R. Arita, A. Yamasaki, K. Held, J. Matsuno, and K. Kuroki, Phys. Rev. B \textbf{75,} 174521 (2007).
\bibitem{Nianalog1} V. I. Anisimov, D. Bukhvalov, and T. M. Rice, Phys. Rev. B \textbf{59,} 7901 (1999).
\bibitem{osmtLNO1} J. Chaloupka and G. Khaliullin, Phys. Rev. Lett. \textbf{100,} 016404 (2008).
\bibitem{osmtLNO2} P. Hansmann, X. Yang, A. Toschi, G. Khaliullin, O. K. Andersen, and K. Held, Phys. Rev. Lett. \textbf{103,} 016401 (2009).
\bibitem{IrhalfJ1} B. J. Kim, H. Jin, S. J. Moon, J.-Y. Kim, B.-G. Park, C. S. Leem, J. Yu, T. W. Noh, C. Kim, S.-J. Oh, J.-H. Park, V. Durairaj, G. Cao, and E. Rotenberg, Phys. Rev. Lett. \textbf{101,} 076402 (2008).
\bibitem{IrhalfJ2} B. J. Kim, H. Ohsumi, T. Komesu, S. Sakai, T. Morita, H. Takagi, and T. Arima, Science \textbf{323,} 1329 (2009).
\bibitem{Iranalog1} F. Wang and T. Senthil, Phys. Rev. Lett. \textbf{106,} 136402 (2011). 
\bibitem{Iranalog2} R. Arita, J. Kune\v{s}, A. V. Kozhevnikov, A. G. Eguiluz, and M. Imada, arXiv: 1107.0835.
\bibitem{NiJ} K. Nakajima, K. Yamada, S. Hosoya, T. Omata, and Y. Endoh, J. Phys. Soc. Jpn. \textbf{62,} 4438 (1993).
\bibitem{Nipoly1} Y. Takeda, R. Kanno, M. Sakano, O. Yamamoto, M. Takano, Y. Bando, H. Akinaga, K. Takita, and J. B. Goodenough, Mat. Res. Bull. \textbf{25,} 293 (1990).
\bibitem{Nipoly2} R. J. Cava, B. Batlogg, T. T. Palstra, J. J. Krajewski, W. F. Peck, Jr., A. P. Ramirez, and L. W. Rupp, Jr., Phys. Rev. B \textbf{43,} 1229 (1991).
\bibitem{Nithin} S. Shinomori, Y. Okimoto, M. Kawasaki, and Y. Tokura, J. Phys. Soc. Jpn. \textbf{71,} 705 (2002).
\bibitem{Nistripe1} C. H. Chen, S.-W. Cheong, and A. S. Cooper, Phys. Rev. Lett. \textbf{71,} 2461 (1993).
\bibitem{Nistripe2} J. M. Tranquada, D. J. Buttrey, and V. Sachan, Phys. Rev. B \textbf{54,} 12318 (1996).
\bibitem{Nistripe3} H. Yoshizawa, T. Kakeshita, R. Kajimoto, T. Tanabe, T. Katsufuji, and Y. Tokura, Phys. Rev. B \textbf{61,} R854 (2000).
\bibitem{NiCB1} R. Kajimoto, K. Ishizaka, H. Yoshizawa, and Y. Tokura, Phys. Rev. B \textbf{67,} 014511 (2003).
\bibitem{NiCB2} K. Ishizaka, Y. Taguchi, R. Kajimoto, H. Yoshizawa, and Y. Tokura, Phys. Rev. B \textbf{67,} 184418 (2003).
\bibitem{NiFS} M. Uchida, K. Ishizaka, P. Hansmann, Y. Kaneko, Y. Ishida, X. Yang, R. Kumai, A. Toschi, Y. Onose, R. Arita, K. Held, O. K. Andersen, S. Shin, and Y. Tokura, Phys. Rev. Lett. \textbf{106,} 027001 (2011).
\bibitem{Cureview} A. Damascelli, Z. Hussain, and Z.-X. Shen, Rev. Mod. Phys. \textbf{75,} 473 (2003).
\bibitem{BL} H. Ohashi, Y. Senba, H. Kishimoto, T. Miura, E. Ishiguro, T. Takeuchi, M. Oura, K. Shirasawa, T. Tanaka, M. Takeuchi, K. Takeshita, S. Goto, S. Takahashi, H. Aoyagi, M. Sano, Y. Furukawa, T. Ohata, T. Matsushita, Y. Ishizawa, S. Taniguchi, Y. Asano, Y. Harada, T. Tokushima, K. Horiba, H. Kitamura, T. Ishikawa, and S. Shin, AIP Conf. Proc. \textbf{879,} 523 (2007).
\bibitem{LMTO} O. K. Andersen and T. Saha-Dasgupta, Phys. Rev. B \textbf{62,} R16219 (2000).
\bibitem{selectionrule} S. H\"{u}fner, \textit{Photoelectron Spectroscopy} (Springer, Berlin/Heidelberg/New York, 2003).
\bibitem{OKA} O. K. Andersen, A. I. Liechtenstein, O. Jepsen, and F. Paulsen, J. Phys. Chem. Solids \textbf{56,} 1573 (1995).
\bibitem{firstgiant} F. Ronning, K. M. Shen, N. P. Armitage, A. Damascelli, D. H. Lu, Z.-X. Shen, L. L. Miller, and C. Kim, Phys. Rev. B \textbf{71,} 094518 (2005).
\bibitem{hierar} W. Meevasana, X. J. Zhou, S. Sahrakorpi, W. S. Lee, W. L. Yang, K. Tanaka, N. Mannella, T. Yoshida, D. H. Lu, Y. L. Chen, R. H. He, H. Lin, S. Komiya, Y. Ando, F. Zhou, W. X. Ti, J. W. Xiong, Z. X. Zhao, T. Sasagawa, T. Kakeshita, K. Fujita, S. Uchida, H. Eisaki, A. Fujimori, Z. Hussain, R. S. Markiewicz, A. Bansil, N. Nagaosa, J. Zaanen, T. P. Devereaux, and Z.-X. Shen, Phys. Rev. B \textbf{75,} 174506 (2007).
\bibitem{giantboson} T. Valla, T. E. Kidd, W.-G. Yin, G. D. Gu, P. D. Johnson, Z.-H. Pan, and A. V. Fedorov, Phys. Rev. Lett. \textbf{98,} 167003 (2007).
\bibitem{giantNCCO} F. Schmitt, B. Moritz, S. Johnston, S.-K. Mo, M. Hashimoto, R. G. Moore, D.-H. Lu, E. Motoyama, M. Greven, T. P. Devereaux, and Z.-X. Shen, Phys. Rev. B \textbf{83,} 195123 (2011).
\bibitem{pgBitwo} Y. Kohsaka, C. Taylor, P. Wahl, A. Schmidt, J. Lee, K. Fujita, J. W. Alldredge, K. McElroy, J. Lee, H. Eisaki, S. Uchida, D.-H. Lee, and J. C. Davis, Nature \textbf{454,} 1072 (2008).
\bibitem{Ramantwo} M. Le Tacon, A. Sacuto, A. Georges, G. Kotliar, Y. Gallais, D. Colson, and A. Forget, Nature Phys. \textbf{2,} 537 (2006).
\bibitem{LSCOtwo} T. Yoshida, M. Hashimoto, S. Ideta, A. Fujimori, K. Tanaka, N. Mannella, Z. Hussain, Z.-X. Shen, M. Kubota, K. Ono, S. Komiya, Y. Ando, H. Eisaki, and S. Uchida, Phys. Rev. Lett. \textbf{103,} 037004 (2009).
%\newpage
\end{thebibliography}
\end{document}